**Gate-induced insulator to band-like transport transition in organolead halide perovskite**

*Dehui Li, Hung-Chieh Cheng, Hao Wu, Yiliu Wang, Jian Guo, Gongming Wang, Yu Huang, and Xiangfeng Duan*[*]

Dr. Dehui Li, Yiliu Wang
Department of Chemistry and Biochemistry, University of California, Los Angeles, California 90095, USA
Hung-Chieh Cheng, Hao Wu, Jian Guo
Department of Materials Science and Engineering, University of California, Los Angeles, California 90095, USA.
Dr. Gongming Wang, Prof. Xiangfeng Duan
Department of Chemistry and Biochemistry and California Nanosystems Institute, University of California, Los Angeles, California 90095, USA
Email: xduan@chem.ucla.edu;
Prof. Yu Huang
Department of Materials Science and Engineering and California Nanosystems Institute, University of California, Los Angeles, California 90095, USA.

The organolead halide perovskite materials have recently attracted intensive interest as a most promising material for high efficiency and low-cost solution processable optoelectronic applications.[1-7] In particular, due to the high optical absorption coefficient, moderate charge mobility and very long diffusion length, the power conversion efficiency of the organolead halide perovskite solar cells has soared to a certified efficiency of more than 20% within past several years.[3, 8] In addition to its application in the field of solar cells, the organolead halide perovskite based light-emitting devices,[9] lasers[5, 10, 11] and photodetectors[12] have also been demonstrated with fairly decent performance. Understanding the charge transport mechanism in organolead halide perovskite materials is essential for further improving the performance of these devices and designing new device architectures.

However, despite the rapid advancement in the perovskite-based optoelectronic applications, the studies on the charge transport in organolead halide perovskite is still in its infancy stage and largely limited to spin-coated perovskite thin films due to its solubility of organolead halide perovskite materials in many common solvents such as water, acetone and isopropanol, incompatibility with typical lithographic device fabrication steps and instability in the ambient condition.[13-15] Previous studies demonstrated that the charge carrier mobility is modest due to the scattering because the carriers in organolead halide perovskite have low effective mass.[16-18] The trap densities are rather low and thus the impurity scattering cannot be the dominant factor that limits the charge carrier mobility in the organolead halide perovskite.[16, 19, 20] Charge carrier-phonon scattering is believed to be the primary scattering mechanism which has been demonstrated by several groups via time-resolved carrier dynamics, charge transport measurement and Hall measurement.[21-24] *X. Y. Zhu* and *V. Podzorov* proposed that the charge carriers in organolead halide perovskite might be protected as large polarons in the tetragonal phase with a large charge carrier effective mass based on a series of experimental facts including the long carrier diffusion length, long lifetime, low electron-hole recombination rate, low scattering rate and temperature dependent mobility[25-27], while *A. J. Neukirch et al.* reported the formation of small polarons in the organolead halide perovskites via in-depth computation.[28] The large polaron model predicts a large charge carrier effective mass and is supported by the large difference between the static and high-frequency dielectric constants,[25, 26] which contradicts the low charge carrier effective mass predicted by first-principle calculation[17] and experimentally determined by magneto-absorption.[17, 18] In terms of the transport within orthorhombic phase, the dielectric constant becomes rather small, which might lead to the free carrier transport. Overall, the charge carrier transport mechanism remains elusive and an actively debated topic that demands for further investigation. Here we report a gate-induced insulator-to-band-like transport transition in the orthorhombic phase of the organolead halide perovskite microplate devices by using monolayer graphene as the low-barrier source-drain electrical contact material.

**Figure 1**a displays the schematic of a two-terminal organolead halide perovskite microplate field-effect transistor (FET) with monolayer graphene as the source-drain contacts. Figure 1b exhibits an optical image of such device with the monolayer graphene stripes outlined by red dashed lines. The graphene-contacted devices are fabricated via mechanical exfoliation and dry alignment transfer method (See Experimental Section).[29, 30] All electrical measurements are carried out in dark from room temperature to liquid nitrogen temperature by using a cryogenic probe station. Since the monolayer graphene has a finite density of states, the work function of monolayer graphene can be easily tuned by the back-gate voltage,[29] which opens the possibility to tune the barrier height between the graphene and organolead halide perovskite microplates. In addition, our previous studies indicate that unipolar n-type behavior can be observed in monolayer or bilayer graphene contacted organolead halide perovskite microplate FETs, which could be attributed to the relative small difference between the work function of graphene and the conduction band edge of organolead halide perovskite, in contrast to the Au electrode contact where ambipolar behavior was observed both in spin-coated thin film and microplate devices. The small or negligible contact barrier between the monolayer graphene and organolead halide perovskite leads to the presence of the insulator to band-like transport transition in our FETs.

The output characteristics ($I_{sd}$ vs. $V_{sd}$) of the monolayer graphene-contacted organolead halide perovskite microplate FET at 77 K show that the source-drain current $I_{sd}$ continuously increases with increasing positive gate voltage, indicating an n-type semiconductor behavior (Figure 1c). The transfer characteristics also exhibit a unipolar n-type behavior for the monolayer graphene-contacted device (Figure 1d). The similar unipolar n-type behavior has been observed for bilayer graphene contacted perovskite microplate FETs (**Figure S1**). Nevertheless, the unipolar n-type behavior evolves to ambipolar behavior for Au contacted FETs (Figure S1), which further confirms that the observed unipolar behavior for mono- and bi-layer graphene contacted organolead halide perovskite microplate FETs is due to the smaller difference between the work function of graphene and the conduction band edge of organolead halide perovskite.

The electron field-effect mobility can be extracted from transfer characteristics (Figure 1e). Similar to what observed in spin-coated thin film organolead halide perovskite FETs, our devices show considerable hysteresis in the transfer curves, which prevents the precise evaluation of the mobility and can lead to systematic errors for mobility determination and misinterpretation of temperature dependent source-drain current under different gate voltages. Nevertheless, both positive and negative sweepings give the same trend of mobility and source-drain current versus temperature. In the rest of this report, we focus on the mobility and source-drain current extracted from the positive sweeping for the simplicity of discussion. In general, the mobility continuously increases with the decrease of temperature, but with a sudden fall around 170 K and starts to increase again with further decreasing temperature below 170 K. The sudden fall of mobility around 170 K can be attributed to the tetragonal-to-orthorhombic phase transition,[31] which was further confirmed by the temperature dependent photoluminescence (PL) spectra (Figure 1f). The presence of two emission peaks below 170 K is an evidence of the tetragonal inclusions within orthorhombic phase (Figure 1f), with the lower energy emission peak originates from the tetragonal phase inclusions, and the orthorhombic phase leads to the higher energy emission peak.[31] While the increase of the mobility with the decreasing temperature above 170 K might be ascribed to the reduction of carrier-phonon interaction or due to large polaron transport which can be protected from scattering and trapping,[24] the increase of mobility below 170 K is a result of reduction of electron-phonon interaction and the decrease of the small inclusions of tetragonal phase domains within orthorhombic phase, as demonstrated by low temperature transmission electron microscopy and temperature dependent PL spectra.[31]

The presence of the tetragonal phase introduces extra grain boundary disorders, which might introduce in-gap electronic states and thus significantly alter the charge transport. To investigate how the tetragonal inclusion introduced disorders influences the charge transport, we display the temperature dependent transfer characteristics of a monolayer graphene contacted device below phase transition temperature around 170 K (**Figure 2**a). By varying temperature below 170 K, there is a source-drain current crossover for the positive sweeping near the gate voltage of 60 V: while the source-drain current increases with the decrease of temperature for the gate voltage larger than 60 V, a complete opposite trend of source-drain current was observed for the gate voltage smaller than 60 V (Figure 2a).  In order to view this trend more clearly, we extract the source-drain current against temperature under different gate voltages as shown in Figure 2b. A clear insulator to band-like transport transition was found with increasing gate voltage within orthorhombic phase (below 180 K), while the source-drain current increases with decreasing temperature under all gate voltages investigated in tetragonal phase (above 180 K). This trend is rather different from that of Au contacted perovskite microplate transistors, where the source-drain current always increases with the decreasing temperature within both tetragonal and orthorhombic phase (**Figure S2**). This continuous increase might be attributed to the contact barrier due to the large difference between the Au work function and the conduction band edge of perovskite.

When organolead halide perovskite is in tetragonal phase above 180K, the continuous increase of source-drain current with the decreasing temperature under all gate voltages within tetragonal phase might be due to the fact the charge carriers are protected as large polarons, which has been proposed previously and primarily supported by optical measurement.[16] In contrast to the small polarons where the thermally activated hopping dominates the charge transport, the large polarons with a large charge carrier effective mass can effectively reduce the carrier-phonon and carrier-defect scattering, leading to the increase of the source-drain current with the decreasing temperature.[25, 32]

Below 180 K within the orthorhombic phase, the small inclusions of tetragonal phase would introduce more grain boundaries and thus the in-gap electronic states. As a consequence, the Fermi level locates within the bandgap at low gate voltages, leading to the insulating behavior. As the gate voltage increases, the Fermi level shifts close to or into mobility edge[33], resulting the band-like transport.[34, 35] Similar gate-induced metal to band-like transport transition has been observed in organic polymer transistor and quantum dot films. The source-drain current in the insulating regime for gate voltage around 40 and 50 V can be described by[36]

$$I \propto \exp(-\sqrt{\frac{T_0}{T}})$$

where $T_0$ is a characteristic temperature (Figure 2c). Similar temperature dependent behavior has been identified in polymers and organic semiconductors and can be attributed to variable range hopping transport with Coulomb interaction.[36-38] Hence, the behavior observed in the insulating regime in our devices might be assigned as variable range hopping for gate voltage around 40-50 V, which is resulted from the carrier hopping between randomly distributed localized electronic states. Nevertheless, the number of disorders varies with the temperature, which renders it difficult to precisely determine the hopping mechanism in the insulating regime.

To further confirm our hypothesis that the insulator to band-like transport transition indeed originates from the phase transition induced disorders, we carried out the same electrical measurement for a thinner organolead halide perovskite microplate transistor, which has a lower phase transition temperature.[31] The transfer characteristics of the monolayer graphene-contacted microplate device with a smaller thickness shows the similar unipolar n-type behavior (**Figure 3**a). The very similar mobility versus temperature behavior was observed for the thinner microplate devices and the temperature dependent mobility curve indicates that the orthorhombic-to-tetragonal phase transition occurs at lower temperature (~130 K (Figure 3b), consistent with previous temperature dependent studies. The source-drain current versus temperature under different gate voltages show that a similar insulator to band-like transport transition is present as well in the thinner microplate device right after the orthorhombic-to-tetragonal phase transition, demonstrating that this transition is indeed due to the phase transition induced disorders (Figure 3c).

The observation of the insulator to band-like transport transition in our devices can be attributed to the better crystalline quality of our organolead halide microplates and graphene electrical contact. Previous studies indicate that the defect states are close to the both the valence and the conduction bands with a lower defect density of $10^{10}$-$10^{11}$ cm$^{-3}$ for the spin-coating perovskite films and single crystals.[19, 20] Our field-effect transport measurement exhibits a transition from a unipolar n-type behavior at higher temperature to an ambipolar behavior at lower temperature, which also verifies that the defect density is low and thus the Fermi level significantly shifts with the decreasing temperature.[31] In addition, the difference between the work function of mono- or bi-layer graphene and the conduction band edge of organolead halide perovskite is smaller compared with that of Au contact, leading to a smaller contact barrier. As a result, the intrinsic insulator to band-like transport transition is not smeared by the contact effect and can be observed here.

In summary, we have observed the insulator to band-like transport transition induced by the gate voltage within the orthorhombic phase in organolead halide perovskite microplate devices with monolayer graphene as electrical contacts. This gate-induced insulator to band-like transport transition might be attributed to the disorders introduced by the tetragonal inclusions within the orthorhombic phase. Our findings are not only important to the fundamental research but also have important practical implications in the development of the electronic and optoelectronic devices at low temperatures, which would have applications in the airplanes and satellites.

**Experimental Section**

*Fabrication of field-effect transistors using graphene as contact:* To fabricate the graphene contact devices, graphene strips (as electrodes) were exfoliated on a clean silicon/ silicon oxide (300 nm) substrate, while the lead di-iodine ($PbI_2$) plates were peeled on polymer stack PMMA spun on a silicon wafer. First, the peeled $PbI_2$ plate was aligned and transferred onto the graphene strips and 5 nm Cr/50 nm Au electrodes for graphene stripes were defined by electron-beam lithography and followed by thermal evaporation and lift-off. Finally, the $PbI_2$ plate was converted to organolead halide perovskite ($CH_3NH_3PbI_3$) by using the vapor phase intercalation.[30]

*Electrical measurements and optical characterizations:* Temperature dependent FET device measurements were carried out in a probe station ((Lakeshore, TTP4) coupled with a precision source/measurement unit (Agilent B2902A). The scanning rate for the transport measurement is 20 V/s and the devices were pre-biased at the opposite voltage for 30 s before each measurement. The PL measurement was conducted under a confocal micro-Raman system (Horiba LABHR) equipped with a 600 gr/mm grating in a backscattering configuration excited by a solid state laser (633 nm) with a power of 3 µW. For the low temperature measurement, a liquid nitrogen continuous flow cryostat (Cryo Industry of America) was used to control the temperature varying from 77 K to 300 K.

**Supporting Information**
Supporting Information is available from the Wiley Online Library or from the author.

**Acknowledgements**
We acknowledge the Nanoelectronics Research Facility (NRF) for technical support. X. D. and Y. H. acknowledge the support from the U.S. Department of Energy, Office of Basic Energy Sciences, Division of Materials Science and Engineering through Award DE-SC0008055.

Received: ((will be filled in by the editorial staff))
Revised: ((will be filled in by the editorial staff))
Published online: ((will be filled in by the editorial staff))

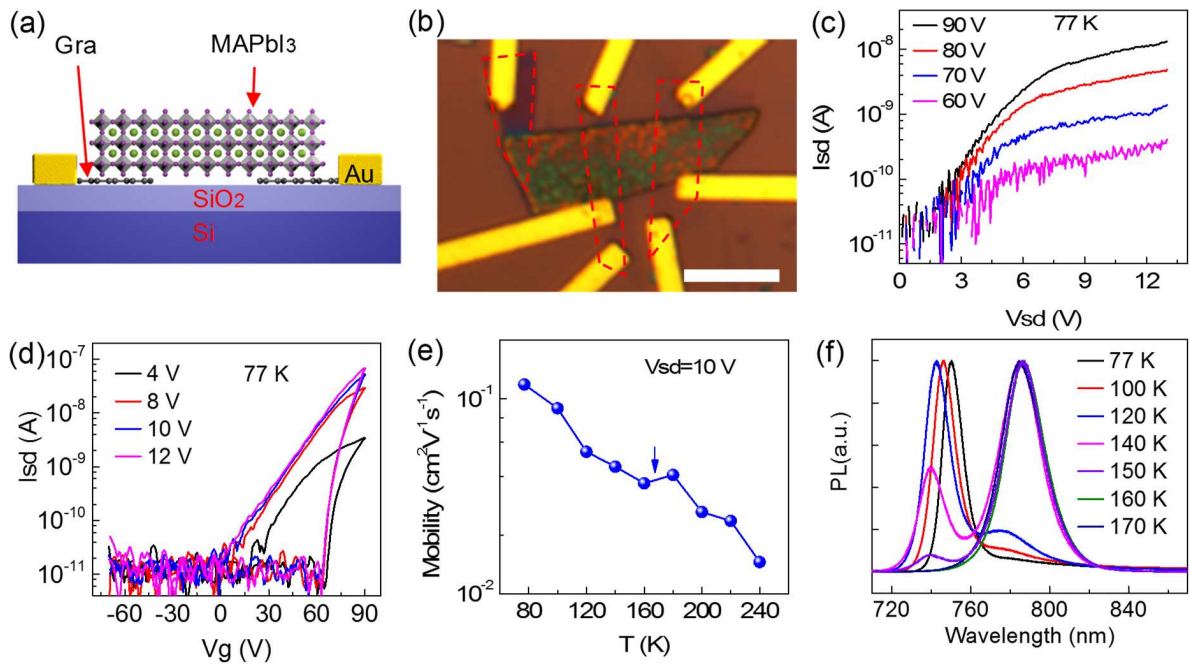

**Figure 1.** (a) Schematic of the back-gate, back-contact halide perovskite microplate field-effect transistor fabricated on a 300 nm $SiO_2$/Si substrate with monolayer graphene stripes as contact. (b) An optical image of the graphene contact device with graphene outlined by red dashed lines. The scale bar is 6 μm. (c-d) The output (Vg= 60, 70, 80, 90 V) (c) and transfer (Vsd=4, 8, 10, 12 V) (d) characteristics of the graphene contact field-effect transistor at 77 K. (e) The temperature dependent field-effect mobility measured with a source-drain voltage of 10 V. The blue arrow indicates the tetragonal-to-orthorhombic phase transition temperature. (f) Temperature dependent photoluminescence spectra for the same halide perovskite microplate under the excitation of a 633 nm laser with a power of 3 μW.

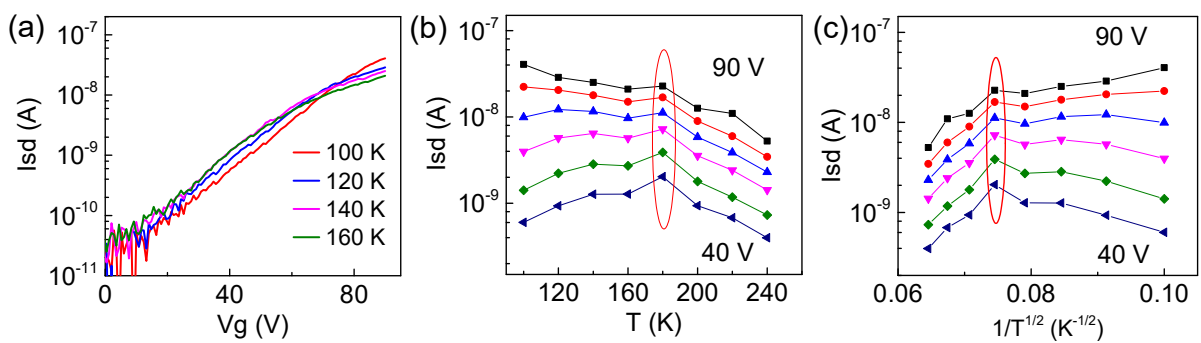

**Figure 2**. (a) The temperature dependent transfer characteristics of the same monolayer graphene-contacted field-effect transistor for positive sweeping. The source-drain voltage is 10 V. (b, c) The temperature dependent source-drain current under different gates voltage for the linear scale (b) and inverse square root scale (c) *x*-axis.

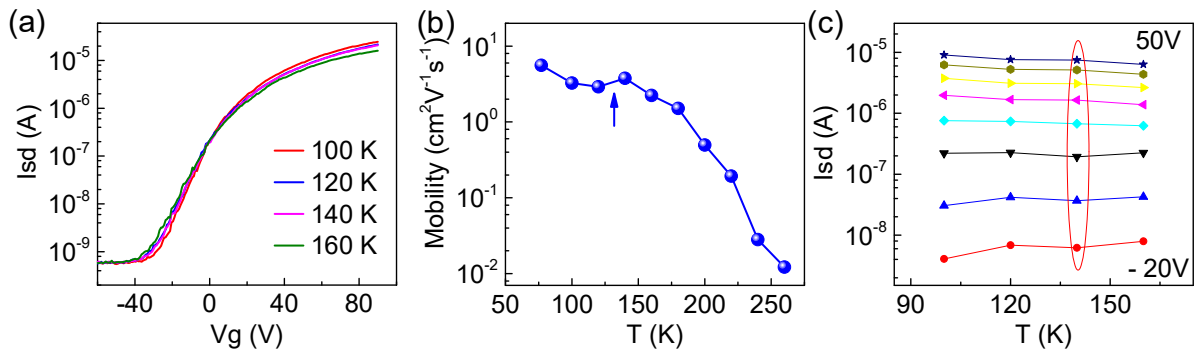

**Figure 3**. (a) The temperature dependent transfer characteristics of the monolayer graphene contacted field-effect transistor for a thinner organolead halide microplate device (positive sweeping). The source-drain voltage is 10 V. The channel length is 10 μm and length is 3 μm. (b) The temperature dependent field-effect mobility measured under a source-drain voltage of 10 V. The blue arrow indicates the tetragonal-to-orthorhombic phase transition temperature. (c) The temperature dependent source-drain current under different gate voltages.

**The table of contents (50-60 words)**

we report the fabrication of organolead halide perovskite microplates with monolayer graphene as low barrier electrical contact. A systematic charge transport studies reveal an insulator to band-like transport transition. Our studies indicate that the insulator to band-like transport transition depends on the orthorhombic-to-tetragonal phase transition temperature and defect densities of the organolead halide perovskite microplates.

Dehui Li, Hung-Chieh Cheng, Hao Wu, Yiliu Wang, Jian Guo, Gongming Wang, Yu Huang, and Xiangfeng Duan[*]

**Title: Gate-induced insulator to band-like transport transition in organolead halide perovskite**

ToC figure

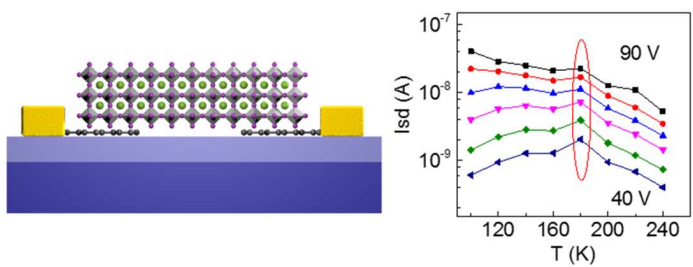

## Supporting Information

**Gate-induced insulator to band-like transport transition in organolead halide perovskite**

Dehui Li, Hung-Chieh Cheng, Hao Wu, Yiliu Wang, Jian Guo, Gongming Wang, Yu Huang, and Xiangfeng Duan[*]

Dr. Dehui Li, Yiliu Wang
Department of Chemistry and Biochemistry, University of California, Los Angeles, California 90095, USA
Hung-Chieh Cheng, Hao Wu, Jian Guo
Department of Materials Science and Engineering, University of California, Los Angeles, California 90095, USA.
Dr. Gongming Wang, Prof. Xiangfeng Duan
Department of Chemistry and Biochemistry and California Nanosystems Institute, University of California, Los Angeles, California 90095, USA
Email: xduan@chem.ucla.edu;
Prof. Yu Huang
Department of Materials Science and Engineering and California Nanosystems Institute, University of California, Los Angeles, California 90095, USA.

Keywords: perovskite, graphene contact, insulating behavior, band-like transport

**The transfer characteristics of the individual organolead halide perovskite microplate field-effect transitions with different electrical contact**

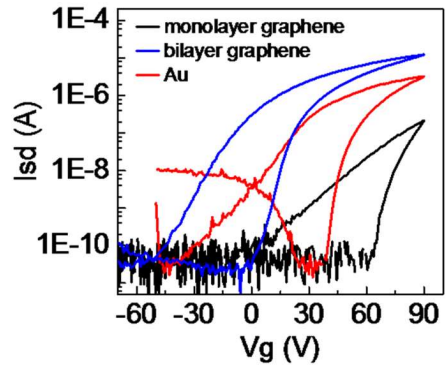

**Figure S1** The transfer characteristics of the individual organolead halide perovskite microplate field-effect transitions with different electrical contact at 77 K. The source-drain voltage is 10 V.

**The temperature dependent source-drain current under different gate voltages for Au-contacted devices**

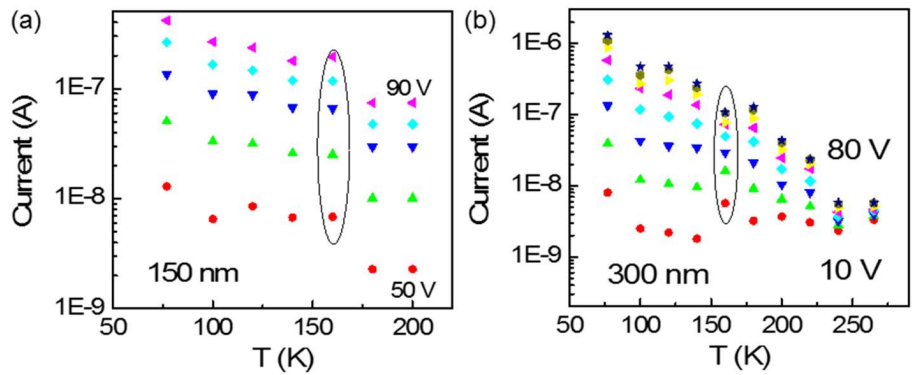

**Figure S2**. The temperature dependent source-drain current under different gate voltages for 150 nm (a) and 300 nm (b) thick perovskite microplate transistors with Ti/Au as electrical contact materials.